\documentclass[preprint,floatfix] {revtex4} 
\newcommand{\rvec}{\mathrm {\mathbf {r}}} 
\newcommand{\pvec}{\mathrm {\mathbf {p}}} 
\usepackage{graphicx}
\usepackage{subfigure}
\begin{document}

\title{Ground and excited states of spherically symmetric potentials through an imaginary-time evolution method:
Application to spiked harmonic oscillators}
\author{Amlan K. Roy}
\altaffiliation{Email: akroy@iiserkol.ac.in, akroy6k@gmail.com}
\affiliation{Division of Chemical Sciences,   
Indian Institute of Science Education and Research Kolkata, 
Mohanpur Campus, Nadia, 741252, WB, India.}

\begin{abstract}
Starting from a time-dependent Schr\"odinger equation, stationary states of 3D central potentials are obtained. An imaginary-time evolution
technique coupled with the minimization of energy expectation value, subject to the orthogonality constraint leads to ground and excited
states. The desired diffusion equation is solved by means of a finite-difference approach to produce accurate wave functions, energies, 
probability densities and other expectation values. Applications in case of 3D isotropic harmonic oscillator, Morse as well 
the spiked harmonic oscillator are made. Comparison with literature data reveals that this is able to produce high-quality and 
competitive results. The method could be useful for this and other similar potentials of interest in quantum mechanics. 
Future and outlook of the method is briefly discussed. 

\end{abstract}
\maketitle

\section{Introduction}
Applications of quantum mechanics in various branches of physics, chemistry, biology, such as atomic, 
molecular, nuclear physics, particle physics, astrophysics, etc., often require solution of 
Schr\"odinger equation (SE). The system is characterized by an external potential term present in the Hamiltonian 
operator. Leaving aside a few occasions, such as the well-known harmonic oscillator or Coulomb potential
representing some idealized situations, \emph{exact} analytical solution in most of these problems remains 
elusive. Search for such solutions is appealing and have been pursued by a large number of researchers.
In recent years, such analytical solutions have been reported for few more potentials such as 
Kratzer-Fues potential in N dimension \cite{oyewumi05}, Mie \cite{sever08}, pseudoharmonic potential
in 2D \cite{dong02a}, 3D \cite{sever08a} and N-dimension \cite{wang02}, Morse \cite{dong02b}, P\"oschl-Teller 
\cite{dong02c}, Manning-Rosen \cite{dong07a,qiang07} and some other diatomic molecular potentials \cite{akcay12}.
Nevertheless they are few and far between,
and it is imperative that alternate approximation methods be developed. Therefore, a variety of accurate, 
efficient, elegant methodologies for such solutions have been put forth over the years.
This encompasses a wide range of analytic, semi-analytic and numerical techniques. The 
literature is vast; here we refer to some of the most prominent ones, \emph{viz.}, Nikiforov-Uvarov method 
\cite{nikiforov88}, super-symmetric quantum mechanics \cite{fellows09} asymptotic iteration method 
\cite{ciftci03,ciftci05}, exact quantization rule \cite{ma05,ma05a}, factorization method \cite{dong07}, 
wave function ansatz approach \cite{dong01,dong02}, generalized pseudospectral method \cite{roy04,roy04a}, 
proper quantization rule \cite{qiang10,serrano10}, etc.

In all the above mentioned approaches, approximate solutions are obtained starting from a time-independent
SE (TISE). In this work, we explore the possibility of an approximate solution based on time-dependent
Schr\"odinger equation (TDSE) instead. This is achieved by transforming the TDSE in imaginary time to a diffusion
equation, followed by a minimization of the energy expectation value to reach the global minimum. Such a 
technique was adopted in connection with a 
random-walk simulation of the solution of \emph{ab initio} SE for electronic systems such as, H $^2P$, H$_3^+$ 
(D$_{3h}$) $^1A_1$, H$_2$ $^3\Sigma^+_u$, H$_4$ $^1\Sigma^+_g$, Be $^1S$, CH$_4$, etc., 
\cite{anderson75, anderson76, garmer87}. In a separate work, eigenvalues, eigenfunctions of TDSE were obtained by evolving 
the same in imaginary time and representing the Hamiltonian in a grid by a relaxation method \cite{kosloff86}. 
Representative applications were given for Morse potential, H\'enon-Heiles 
system and weakly bound states of He on a Pt surface. Another interesting route (the so-called spectral method), based on the 
grid, to exploit TDSE for obtaining eigenvalues, eigenfunctions was adopted in \cite{feit82}, whereby the initial wave function 
was propagated for long time. 
Then eigenvalues are obtained by performing a Fourier transform of the auto-correlation function of propagated wave with the 
initial wave function. In yet another development, imaginary-time evolution technique 
was applied for direct calculation of \emph{ground-state} densities and other properties of noble gas atoms, ions 
such as He, Be$^{++}$, Ne, Ar, Kr, Xe, as well as molecules like H$_2$, HeH$^+$, He$_2^{++}$, from 
the solution of a single \emph{TD quantum fluid dynamical} equation of motion \cite{dey99,roy99,roy02}. Later,                   
ground as well as excited-state energies, densities and other expectation values of 1D anharmonic and double-well \cite{roy02a},
multiple-well \cite{gupta02} and self-interacting nonlinear \cite{wadehra03} oscillators were obtained with
impressive accuracy through this route. Extension was made to ground and low-lying excited 
states of double well potentials in 2D \cite{roy05}. Furthermore, during the same time period, a finite-difference time 
domain approach was suggested for solution of the respective TDSE in imaginary time. Applications were made to the problems of 
infinite square potential, quantum anharmonic oscillators in 1D, 2D, 3D, as well as hydrogen atom 
\cite{sudiarta07} with reasonable success. Later, this scheme was employed for a charged particle in magnetic 
field \cite{sudiarta08}, and for the computation of thermal density matrix of a single-particle confined 
quantum system \cite{sudiarta09}. An optimized parallelization scheme for solving 3D SE has been presented lately \cite{strickland10}. 
Imaginary-time propagation technique has also been exploited in numerical solution
of eigenvalues, eigenfunctions of large matrices originating from discretization of linear and non-linear
SE by means of split-operator method \cite{lehtovaara07}, and also for large-scale 2D eigenvalue problems in presence of a 
magnetic field \cite{luukko13}. Improved, high-order, imaginary-time propagators for 3D SE has been proposed \cite{chin97,chin09}, 
as well as a fourth order algorithm for solving local SE in a homogeneous magnetic field \cite{aichinger05}. 

The purpose of this communication is to present and explore the feasibility of the imaginary time evolution method, as implemented
in \cite{dey99,roy99,roy02,roy02a,gupta02,wadehra03,roy05}, in the context of spherically symmetric 3D potentials. 
As mentioned above, while for the atoms and molecules in first three references, the diffusion equation originated from an 
amalgamation of quantum fluid dynamics and density functional theory, which eventually lead to a TD generalized non-linear SE, 
in latter four references the same arose from the TDSE instead. The current work concern the latter. It is worthwhile mentioning
that while ground and excited states were treated in 1D and 2D using this approach \cite{roy02a,gupta02,wadehra03,roy05}, for 
spherically symmetric potentials (such as for atoms, as in
\cite{dey99,roy99,roy02}, only \emph{ground} states were attempted. Here we include excited states in our study, that can 
extend its domain of applicability to some other systems. After making 
some experiments on pedagogical cases like 3D isotropic quantum harmonic oscillator and Morse potential, we focus on the specific case of 
spiked harmonic oscillator, for illustration. Because of their many interesting properties and considerable challenges they pose, 
these have been investigated by a number of workers in the past three decades varying in their complexity and accuracy. In 
particular, we focus on the eigenvalues, position expectation values, radial densities of ground and low-lying excited
states. At present, we consider the \emph{non-rotational} $(\ell =0)$ case, while \emph{rotational} $(\ell \neq 0)$ situations 
may be studied in future works. The article is organized as follows. In Section II, we give an outline of the 
theoretical framework and details of numerical implementation. Obtained results are discussed in Section III along with a 
comparison with literature results. Finally a few concluding remarks are made in Section V. 

\section{The methodology and numerical implementation}
The TDSE of a single particle moving in a time-independent potential field $V(\rvec)$ is:
\begin{equation}
i \frac{\partial}{\partial t} \psi (\rvec, t) = H \psi(\rvec, t) = 
\left[ -\frac{1}{2} \nabla^2 + V(\rvec) \right] \psi(\rvec, t), 
\end{equation}
where H is the Hamiltonian operator consisting of kinetic and potential energy components. Here and what follows, 
we employ atomic units. The general solution can be expanded in terms of a set of eigenfunctions $\{ \phi_k\}$ and expansion 
coefficients $\{C_k\}$ as follows:
\begin{equation}
\psi(\rvec, t) = \sum_{k=0}^{\infty} C_k \phi_k(\rvec) \exp{(-i\epsilon_kt)}.
\end{equation}
The eigenfunctions $\phi_k(\rvec)$ and eigenvalues $\epsilon_k$ are obtained from the associated 
TISE. Following previous works (see, for example, \cite{dey99,roy99,roy02, roy02a, 
gupta02, wadehra03, roy05}, and references therein), we assume the validity of TDSE in imaginary time $\tau$, and write Eq.~(1) 
in $\tau$. Further, replacing $\tau$ by $-it$, where $t$ is real time, yields (for brevity, $\psi(\rvec,t)$ now refers to the
diffusion function), 
\begin{equation}
-\frac{\partial \psi(\rvec, t) }{\partial t} = H \psi(\rvec, t)
\end{equation}
Defining the time differential operator, $D_t = \frac{\partial}{\partial t}$, transforms this into a nonlinear diffusion-like 
equation, which resembles a diffusion-quantum Monte Carlo equation \cite{hammond94}, 
\begin{equation}
-D_t \ \psi(\rvec, t) = H \psi(\rvec, t).
\end{equation}
One may express $\psi(\rvec, t)$ as, 
\begin{equation} 
\psi(\rvec, t) = C_0 \phi_0 (\rvec)+ \sum_{k=1}^{\infty} C_k \ \phi_k(\rvec) \ e^{-(\epsilon_k-\epsilon_0)t},
\end{equation}
where $\phi_0$, $\epsilon_0$ refer to ground-state eigenfunction and eigenvalue. Hence, taking due account of normalization, one 
observes that, at $t \rightarrow \infty$, $\psi(\rvec,t) \rightarrow \phi_0$, i.e.,  
\begin{equation}
\lim_{t \rightarrow \infty} \psi(\rvec, t) \approx C_0 \phi_0 (\rvec).
\end{equation}
This implies that at any finite non-zero time, $\psi(\rvec, t)$ can be seen as a linear combination of TISE eigenfunctions $\{\phi_k\}$
with appropriate TD coefficients $\{C_k\}$ which decay exponentially in real time. Therefore, numerically propagating $\psi(\rvec,t)$ to a 
sufficiently long time leads to the stationary ground-state wave function (provided $C_0 \neq 0$), apart from a normalization constant, 
corresponding to the global minimum value of $\langle \psi(\rvec,t) |H| \psi(\rvec,t) \rangle$. This is a general technique for solving 
stationary-state eigenvalue problems in quantum mechanics. 

Now let us consider the numerical solution of Eq.~(4). In order to accomplish the time propagation of 
$\psi(\rvec, t)$, one can use a Taylor expansion of $\psi(\rvec, t+\Delta t)$ around time $t$,   
\begin{equation}
\psi(\rvec, t+\Delta t)= \left[ 1+ \Delta t D_t + \frac{(\Delta t)^2}{2!}D_t^2+ \cdots \right] \psi(\rvec, t)
= e^{\Delta t D_t} \psi(\rvec, t).
\end{equation}
From Eq.~(4), we see that $H=-D_t$. Hence the above equation can be rewritten as, 
\begin{equation}
\psi(\rvec, t+\Delta t)= e^{-\Delta tH} \psi(\rvec, t).
\end{equation}
The time-propagator $e^{-\Delta t H}$ is an evolution operator advancing the diffusion function 
$\psi(\rvec,t)$ from an initial time $t$ to next time level $\psi(\rvec, t+\Delta t)$. It is worth mentioning that, this is a real, 
non-unitary operator; hence normalization of $\psi(\rvec, t)$ at an arbitrary time $t$ does not automatically guarantee normalization of 
$\psi(\rvec, t+\Delta t)$ at a future time, $t+\Delta t$. 

At this point, we focus on the important case of \emph{central} force, which is derived from a potential energy function that is 
spherically symmetric, i.e., $V(\rvec)=V(r)$. For this, we discretize the radial variable $r$ (of spherical polar coordinates) according 
to the following, 
\begin{eqnarray}
r_j  & = & x_j^2   \\
x_j  & = & \delta + j \Delta x =\delta +jh,   \ \ \ \ \ \ \   j=1,2,3,\cdots, N.  \nonumber 
\end{eqnarray}
Here $\Delta x = h$ denotes grid spacing in radial coordinate, $\delta$ is a small number ($10^{-6}$ a.u., in present case), 
integer $j$ signifies the counter of increment in spatial direction, while $N$ is the total number of radial points. The operator 
$H$, given in spherical polar coordinates, by, 
\begin{equation}
H_r= - \frac{1}{2} \frac{d^2}{dr^2} - \frac{1}{r} \frac{d}{dr} + \left[ \frac{\ell(\ell+1)}{2r^2} +V(r) \right]
    =- \frac{1}{2} \frac{d^2}{dr^2} - \frac{1}{r} \frac{d}{dr} + v_{\mathrm{eff}}(r) 
\end{equation}
can be recast in transformed $x$-grid as below (terms in the parenthesis correspond to $v_{\mathrm{eff}} (r)$),
\begin{equation}
H= -\frac{1}{8x^2} D_x^2 - \frac{3}{8x^3} D_x + v_{\mathrm{eff}}(r) = aD_x^2 + bD_x + v_{\mathrm{eff}}(r).
\end{equation}
In the above equation, $a=-\frac{1}{8x^2}$, $b=-\frac{3}{8x^3}$, $\ell$ signifies the angular momentum quantum number, while 
$D_x = \frac{d}{dx}$, $D_x^2= \frac{d^2}{dx^2}$ denote 1st, 2nd partial spatial derivatives. Such a radial grid has been 
found to be quite effective and successful for Coulombic 
systems \cite{dey99,roy99}, for it provides a finer grid at small $r$ and coarser grid at large $r$. Subscripts in 
derivatives emphasize that these quantities are given in transformed grid $x$. 
In such a grid, Eq.~(8) can be expressed as ($j,n$ identify the increments in space and time coordinates respectively), 
\begin{equation}
\psi_j^{'(n+1)}= e^{-\Delta tH_j} \ \psi_j^n,
\end{equation}
where a prime signifies an unnormalized diffusion function. This equation can be further written in an equivalent symmetric 
form, given below, 
\begin{equation}
e^{(\Delta t/2) H_j} \ \psi_j^{'(n+1)}= e^{-(\Delta t/2)H_j} \ \psi_j^n.
\end{equation}
By making use of Eq.~(11), above equation can be further recast into a form, as below,
\begin{equation}
e^{(\Delta t/2) (aD_x^2+bD_x+v_{\mathrm{eff}})} \ \psi_j^{'(n+1)} = e^{-(\Delta t/2) (aD_x^2+bD_x+v_{\mathrm{eff}})} \ \psi_j^n.
\end{equation}
Finally, expanding the exponentials, truncating them after second terms, followed by an approximation of $D_x$ and
$D_x^2$ by two- and three-point difference formulas as below,
\begin{eqnarray}
D_x \ \psi_j^n & \approx & \frac{\psi_{j+1}^n-\psi_{j-1}^n}{\Delta x}, \\ 
D_x^2 \ \psi_j^n & \approx & \frac{\psi_{j-1}^n- 2 \psi_j^n+ \psi_{j+1}^n}{2 (\Delta x)^2},   \nonumber
\end{eqnarray}
a set of $N$ simultaneous equations are obtained as follows:
\begin{equation}
\alpha_j \psi_{j-1}^{'(n+1)} + \beta_j \psi_j^{'(n+1)} + \gamma_j \psi_{j+1}^{'(n+1)} =\xi_j^n.  
\end{equation}
where the quantities $\alpha_j, \beta_j, \gamma_j, \zeta_j^n$ are identified as, 
\begin{eqnarray}
\alpha_j & = & -\frac{\Delta t}{16x_j^2 h^2} + \frac{3\Delta t}{32 x_j^3 h}, \ \ \ 
\beta_j    =   1+\frac{\Delta t}{8x_j^2 h^2} + \frac{\Delta t}{2} v_{\mathrm{eff}}, \ \ \  
\gamma_j  =  -\frac{\Delta t}{16x_j^2 h^2} - \frac{3\Delta t}{32 x_j^3 h},  \\
\zeta_j^n & = & \left( \frac{\Delta t}{16x_j^2 h^2} - \frac{3\Delta t}{32 x_j^3 h} \right) \psi_{j-1}^n +
                \left( 1-\frac{\Delta t}{8x_j^2 h^2} - \frac{\Delta t}{2} v_{\mathrm{eff}} \right) \psi_j^n +
                \left( \frac{\Delta t}{16x_j^2 h^2} + \frac{3\Delta t}{32 x_j^3 h} \right) \psi_{j+1}^n .  \nonumber 
\end{eqnarray}
Note that since discretization and truncation occurs on both sides of Eq.~(14), cancellation of error may
occur. Here, $\psi_{j-1}^{'(n+1)}$, $\psi_j^{'(n+1)}$, $\psi_{j+1}^{'(n+1}$ denote the unnormalized diffusion 
functions at time $t_{n+1}$ at radial grids $x_{j-1}, x_j, x_{j+1}$ respectively. The quantities $\alpha_j$, 
$\beta_j$, $\gamma_j$ and $\xi_j^n$ are identical to those appearing in \cite{roy99} except the obvious differences in 
$v_{\mathrm{eff}}$. For the sake of completeness, however, we provide them here. As can be seen, these are expressed in terms of 
$x_j^2$, $x_j^3$, as well as the space and time 
spacings $\Delta x$, $\Delta t$, while $v_{\mathrm{eff}}$ entering in $\beta_j$ and $\xi_j^n$ only.
Also note that $\xi_j^n$ requires knowledge of $\psi_{j-1}^n, \psi_j^n, \psi_{j+1}^n$, the normalized diffusion
functions at radial grids $x_{j-1}, x_j$ and $x_{j+1}$ respectively at time step $t_n$. Equation (16) may further
be rewritten in a convenient, tridiagonal matrix form,
\begin{equation}
\left[ \begin{array}{cccccc}
\beta_1  &  \gamma_1  &          &        &           &    (0)       \\
\alpha_2 &  \beta_2   & \gamma_2 &        &           &              \\
         &  \ddots    & \ddots   & \ddots &           &              \\
         &            & \ddots   & \ddots & \ddots    & \gamma_{N-1} \\
   (0)   &            &          &        & \alpha_N  & \beta_N 
\end{array} \right]  
\left[ \begin{array}{c}
\psi_1^{'(n+1)} \\ 
\psi_2^{'(n+1)} \\ 
\vdots \\
\psi_{N-1}^{'(n+1)} \\ 
\psi_N^{'(n+1)} 
\end{array} \right]
=
\left[ \begin{array}{c}
\xi_1^n \\ 
\xi_2^n \\ 
\vdots \\
\xi_{N-1}^n \\ 
\xi_N^n 
\end{array} \right].
\end{equation}
This can be efficiently solved for $\{ \psi_j^{'(n+1)} \}$ by using a modified Thomas algorithm \cite{press07}. 

Overall procedure of the calculation then involves following sequence of steps. At time step $n=0$, an 
initial guess of the wave function $\psi_j^0$ is made for all $j$. This is then propagated in accordance with Eq.~(8)
following the procedure described above to obtain $\psi^{'(n+1)}$ at $(n+1)$th time step. At each time step, the wave function
becomes smaller as $r$ assumes large values and finally tends to zero as $r$ goes to infinity. Setting them to zero for large 
$r$ was also found to be equally good provided it covered a sufficiently long radial distance. For an excited state 
calculation, $\psi_j^0$ needs to be orthogonalized to all lower states. Several orthogonalization schemes are available; 
here we have employed the widely used Gram-Schmidt method \cite{greub81}. It is known that, while for smaller number of 
states the method is generally accurate, with increase in number of states, however, this tends to introduce
numerical inaccuracy. Since in present work we are mostly concerned with ground and low-lying states, 
this therefore causes no significant impact on the results obtained. Then $\psi^{'(n+1)}$ is normalized to 
$\psi^{(n+1)}$ and the energy expectation values calculated as $\epsilon_0 = \langle \psi^{(n+1)} | H | \psi^{(n+1)} \rangle$. 
If the difference in energy between two consecutive time steps, 
$\Delta \epsilon = \langle H \rangle^{(n+1)} - \langle H \rangle^n$, drops below a certain prescribed limit, 
then the diffusion function $\psi_j^{(n+1)}$ is stored as the corresponding solution of Hamiltonian $H$. 
Otherwise, $\psi_j^{(n+2)}$ is calculated and above steps repeated until $\Delta \epsilon$
reaches such limit. In this study, a tolerance of $10^{-12}$ was set for energy while 10001 radial grid points were 
used uniformly.  Once the diffusion function reaches the desired convergence in this way, $\psi^{'(n+1)}$ is normalized to 
$\psi^{(n+1)}$, from which the various properties of interest, such as the expectation values, etc., may be obtained as:
$\langle A \rangle^{(n+1)} = \langle \psi^{(n+1)} | A | \psi^{(n+1)} \rangle$.  
Note that, for excited-state calculation, diffusion function must remain orthogonal to all lower states at \emph{all} 
time steps, not just at initial time step. Continuing this procedure, one could then obtain first excited state $\epsilon_1$,
second excited state $\epsilon_2$, third excited state $\epsilon_2$, and so on. 
The grid spacing was adjusted according to the problem, as dictated by nature of the particular potential under study. 
This is mentioned at appropriate places in the discussion that follow. Overlap and energy integrals at each time step 
were evaluated by standard Newton-Cotes quadrature, while finite-difference formulas were used for 
the spatial derivatives \cite{abramowitz64}. 

\begingroup
\squeezetable
\begin{table}
\caption {\label{tab:table1}Calculated eigenvalues, radial expectation values, normalization and 
virial ratios for six lowest states of spherical quantum harmonic oscillator corresponding to $\ell=0$.} 
\begin{ruledtabular}
\begin{tabular}{lllllll}
 Energy\footnotemark[1]  & $\langle r^{-2} \rangle $  & $\langle r^{-1} \rangle$ &  $\langle r^0 \rangle$  &  $\langle r^1 \rangle $ 
         & $\langle r^2 \rangle$      & $\langle V \rangle / \langle T \rangle $    \\
\hline
1.49999999 & 2.000000  & 1.128379  & 1.000000  & 1.128379   &  1.499999  &  0.999999   \\
3.4999999  & 2.000000  & 0.940316  & 1.000000  & 1.692569   &  3.499999  &  0.999999   \\
5.4999999  & 1.99999   & 0.83688   & 1.000000  & 2.11571    &  5.50000   &  1.000000    \\
7.4999999  & 2.00000   & 0.76770   & 0.99999   & 2.46833    &  7.49999   &  0.999999    \\
9.499999   & 1.99998   & 0.71668   & 1.00000   & 2.77689    &  9.49999   &  1.00001      \\
11.499999  & 2.0000    & 0.67678   & 1.00000   & 3.05456    &  11.5000   &  1.00000      \\
\end{tabular}
\begin{tabbing}
$^{\mathrm{a}}${The exact energies \cite{cohen92} of six states are 1.5, 3.5, 5.5, 7.5, 9.5 and 11.5 respectively.}
\end{tabbing}
\end{ruledtabular}
\end{table}
\endgroup

\section{Results and Discussion}
At first, we present some specimen results to test the validity and performance of our method. First one is 
the familiar 3D spherical quantum harmonic oscillator, which is an \emph{exactly solvable} system. Table I
gives 6 lowest states corresponding to rotational quantum number $\ell=0$. At this point it is to be noted that
all results reported in all tables throughout the article are truncated and not rounded-off. Therefore, all
the entries are taken to be correct up to the place they are presented. These calculations are performed in a radial
box size of 10 a.u. Not very exhaustive, but a few sample calculations were made to gauge the variations with respect to grid
parameters. In general, good-quality results could be obtained with even smaller number of points, such as 501 or so,
and accuracy could be improved even further (from reported values) by 2--3 decimal places by increasing number of radial points 
from current values. These are briefly touched upon in a following paragraph. 
However, we have not made any attempt to optimize the grid here, as our primary objective in this work is to 
demonstrate the capability and appropriateness of this method in context of physically important situations. As already known, 
energy levels of isotropic harmonic oscillator 
are given by: $E_{k,\ell}=(k+\ell+\frac{3}{2})=(m+\frac{3}{2})$ a.u., where $k$ is zero or \emph{any even} positive 
integer, $\ell$ can be zero or \emph{any} positive integer, so that $m$ can take on \emph{all} integral values, zero or positive. 
Therefore the two quantum numbers $\ell, m$ must have same parity \cite{cohen92}. We see that the present
results are in excellent agreement with exact values for all states. Additionally, the position expectation values of these 
states in columns 2--6 can also be obtained analytically. We have verified $\langle r^{-2} \rangle$, 
$\langle r^{-1} \rangle$, $\langle r^1 \rangle$ and $\langle r^2 \rangle$ for the first two states. For ground 
state $(k=m=0; \ell=0)$ these are: 2, $\frac{2}{\sqrt{\pi}}$, $\frac{2}{\sqrt{\pi}}$ and $\frac{3}{2}$, while for 
first excited state $(k=m=2; \ell=0)$, these values are 2, $\frac{5}{3\sqrt{\pi}}$, $\frac{3}{\sqrt{\pi}}$, $\frac{15}{2}$ 
respectively. Present calculated values are in good agreement with these estimates. As a further test on quality 
of our eigenfunctions, numerically obtained normalization and virial ratios are also provided in fourth and last columns 
respectively. For the $n$th \emph{stationary} state of a 3D quantum harmonic oscillator, the latter can be obtained from,  
\begin{equation}
\frac{d}{dt} \ \langle \rvec \mathbf{\cdot} \pvec \rangle = \frac{i}{\hbar} \langle [H, \rvec \mathbf{\cdot} \pvec] \rangle 
= 2 \langle T \rangle - \langle \rvec \mathbf{\cdot} \nabla V \rangle = 2 \langle T \rangle - 2 \langle V \rangle = 0,
\end{equation}
so that $\frac{\langle V \rangle} {\langle T \rangle} =1$. This further establishes the reliability and strength 
of our present method. 

\begingroup
\squeezetable
\begin{table}
\caption {\label{tab:table2}Calculated eigenvalues (a.u.) in Morse potential (left panel) and ground states of charged 
harmonic oscillator (right panel) along with literature results. PR signifies Present Result.} 
\begin{ruledtabular}
\begin{tabular}{cllccl}
\multicolumn{3}{c}{Morse oscillator}  &  \multicolumn{3}{c}{Charged harmonic oscillator $(\alpha=1)$}   \\ 
\cline{1-3} \cline{4-6}
$n$  &  Energy (PR)   &  Energy (Reference)  &  $\lambda$  &  Energy (PR) & Energy (Exact\footnotemark[1])  \\
\hline
0      & $-$18.42893218 & $-$18.42893218\footnotemark[2]$^,$\footnotemark[3]$^,$\footnotemark[4]   &
0                           &  1.49999999    &  1.5  \\
1      & $-$8.2867965   & $-$8.2867965\footnotemark[2]$^,$\footnotemark[3]$^,$\footnotemark[4]    &
2                           &  2.499999999   &  2.5  \\
2      & $-$2.1446609   & $-$2.1446609\footnotemark[2]$^,$\footnotemark[3]$^,$\footnotemark[4]    &
$\sqrt{20}$                 &  3.499999999   &  3.5  \\
3      & $-$0.002525     & $-$0.002525\footnotemark[2]$^,$\footnotemark[3]$^,$\footnotemark[4]       &
$\sqrt{30+6\sqrt{17}}$      &  4.499999999   &  4.5  \\
       &                 &                           &
$\sqrt{70+6\sqrt{57}}$      &  5.499999999   &  5.5  \\
       &                 &                           &
14.450001026966             &  6.500000000   &  6.5  \\
       &                 &                           &
18.503131410003             &  7.500000000   &  7.5  \\
\end{tabular}
\begin{tabbing}
$^{\mathrm{a}}${Ref.~\cite{navarro92}. These results have been divided by 2 to take care of a 2 factor.} \\
$^{\mathrm{b}}${Exact result, Ref.~\cite{shore73}.} \hspace{40pt} \=
$^{\mathrm{c}}${B-Splines result, Ref.~\cite{landtman93}.} \hspace{40pt} \=
$^{\mathrm{d}}${Generalized pseudospectral result, Ref.~\cite{roy04a}.} 
\end{tabbing}
\end{ruledtabular}
\end{table}
\endgroup

In Table II, we examine two more special cases where \emph{exact} analytical results are available. First one is the so-called Morse 
potential having following functional form \cite{shore73}:
\begin{equation}
V(r)= 25(e^{-4(r-3)}-2e^{-2(r-3)}), \ \ \ \ \ \ E_n=-\left[ 5-\sqrt{2}(n+\frac{1}{2}) \right]^2, 
\ \ \ \ n=0,1,2,3.
\end{equation}
Morse potential plays a very significant role in the vibration-rotation spectra of diatomic molecules and has been 
extensively studied by a large number of workers ever since its inception about 85 years ago. The above potential
supports only four bound states; corresponding exact analytical energies are given in Eq.~(20) \cite{shore73}. 
In the left panel, our energies for all four states are seen to match exactly with these as well as B-spline result 
\cite{landtman93} and generalized pseudospectral method \cite{roy04a}. It is worth mentioning that for first three 
states the precision of Table II could be reached quite easily with $r_{max}=20$ a.u. only, while same for the fourth state 
requires a value of about 200 a.u. Our second example corresponds to a special case 
of a general class of interaction potentials, known as spiked harmonic oscillators (SHO), characterized by the following 
functional form,
\begin{equation}
V(r)=\frac{1}{2} \left[ r^2+ \frac{\lambda}{r^{\alpha}} \right], \ \ \ \alpha > 0.
\end{equation}
In this equation, coupling parameter $\lambda$ determines strength of perturbative potential, while positive constant $\alpha$ 
defines type of singularity at origin. In a relatively simpler case of $\alpha=1$ (termed as charged harmonic oscillator), the 
system does not exhibit 
super-singularity and the Hamiltonian assumes a simplified confined Coulomb potential type form effectively. It has been 
pointed out that such a system offers an infinite set of \emph{elementary} solutions. The right panel compares seven 
such elementary solutions in ground state of a charged harmonic oscillator along with exact results \cite{navarro94}. 
Note, the first one $(\lambda=0)$ refers to trivial case of an unperturbed Hamiltonian, i.e., a quantum harmonic 
oscillator having energy $E=3/2$. The other $\lambda$'s are taken from solutions of the polynomial equation \cite{navarro94}. 
All these ground states are obtained by engaging a radial grid of 10 a.u. In all these instances, current energies match 
excellently with exact values.

\begingroup
\squeezetable
\begin{table}
\caption {\label{tab:table3}Calculated eigenvalues and selected expectation values (a.u.) of charged harmonic oscillator for several 
positive and negative values of $\lambda$. First six eigenstates corresponding to $\ell=0$ are given. Numbers in the parentheses are 
quoted from Ref.~\cite{roy04}.} 
\begin{ruledtabular}
\begin{tabular}{clll|clll}
$\lambda$ &   Energy & $\langle r^{-1} \rangle$ &  $\langle r^1 \rangle$  &   
$\lambda$ &   Energy & $\langle r^{-1} \rangle$ &  $\langle r^1 \rangle$  \\
\hline
$-$0.001 & 1.4994357(1.4994357)          & 1.12854   & 1.12826   & 0.001    &  1.5005641(1.5005641)  & 1.12822  &   1.12850    \\
         & 3.4995298(3.4995298)          & 0.94038   & 1.69251   &          &  3.5004701(3.5004701)  &  0.94025 &   1.69262    \\
         & 5.4995815(5.4995815)          & 0.83692   & 2.11567   &          &  5.5004184(5.5004184)  &  0.83685 &   2.11575    \\
         & 7.4996161                     & 0.76772   & 2.46830   &          &  7.5003838               &  0.76768 &   2.46836    \\
         & 9.4996416                     & 0.71670   & 2.77685   &          &  9.5003583               &  0.71667 &   2.77689    \\
         & 11.4996616                    & 0.67679   & 3.05454   &          &  11.5003383              &  0.67676 &   3.05458    \\
$-$0.01  & 1.4943542                     & 1.12994   & 1.12720   &  0.01    &  1.5056380              &  1.12682 &   1.12955    \\
         & 3.4952968                     & 0.94093   & 1.69201   &          &  3.5047000              &  0.93970 &   1.69313    \\
         & 5.4958147                     & 0.83723   & 2.11535   &          &  5.5041835               &  0.83654 &   2.11607    \\
         & 7.4961609                     & 0.76793   & 2.46806   &          &  7.5038379               &  0.76748 &   2.46860    \\
         & 9.4964161                     & 0.71684   & 2.77666   &          &  9.5035830               &  0.71652 &   2.77708    \\
         & 11.4966158                    & 0.67690   & 3.05438   &          & 11.5033835               &  0.67665 &   3.05473    \\
$-$0.1   & 1.4431875(1.4431875)          & 1.14420   & 1.11659   &  0.1     & 1.5560334(1.5560334)   &  1.11304 &   1.14008    \\
         & 3.4528298(3.4528298)          & 0.94649   & 1.68698   &          & 3.5468614(3.5468614)   &  0.93414 &  1.69818     \\
         & 5.4580701(5.4580701)          & 0.84031   & 2.11209   &          & 5.5417576(5.5417576)     &  0.83342 &  2.11935     \\
         & 7.4615591                     & 0.76993   & 2.46565   &          & 7.5383286                &  0.76544 &  2.47102     \\
         & 9.4641260                     & 0.71827   & 2.77475   &          & 9.5357939                &  0.71507 &  2.77900     \\
         & 11.4661312                    & 0.67797   & 3.05280   &          & 11.5338082               &  0.67555 &  3.05632     \\
$-1$     & 0.8926027                     & 1.31029   & 1.00677   &  1       & 2.0289385               &  0.99421 &  1.24072     \\
         & 3.0145292                     & 1.00105   & 1.63746   &          & 3.9548368               &  0.87947 &  1.74928     \\
         & 5.0733048                     & 0.86907   & 2.08009   &          & 5.9096008                &  0.80118 &  2.15283     \\
         & 7.1108547                     & 0.78819   & 2.44200   &          & 7.8779872                &  0.74381 &  2.49574     \\
         & 9.1379189                     & 0.73109   & 2.75599   &          & 9.8541170                &  0.69938 &  2.79855     \\
         & 11.1588046                    & 0.68756   & 3.03725   &          & 11.8351717               &  0.66356 &  3.07246     \\
$-$10    & $-$12.440500($-$12.440499)    & 5.02341   & 0.29792   &   10     & 5.2887417(5.2887417)   
         &  0.57934\footnotemark[1] &  1.88860\footnotemark[2]     \\
         & $-$2.4172388($-$2.4172388)    & 1.46634   & 1.04083   &          & 7.0754394(7.0754394)   
         &  0.57219\footnotemark[1] &  2.20351\footnotemark[2]     \\
         & 0.8696992(0.8696992)          & 0.98099   & 1.69717   &          & 8.8981164(8.8981164)     
         &  0.56237\footnotemark[1] &  2.49563\footnotemark[2]     \\
         & 3.4282462                     & 0.82558   & 2.16774   &          & 10.7479670               &  0.55143 &  2.76737     \\
         & 5.7738977                     & 0.74308   & 2.54292   &          & 12.6187932               &  0.54019 &  3.02143     \\
         & 8.0207630                     & 0.68873   & 2.86318   &          & 14.5061493               &  0.52906 &  3.26020     \\
\end{tabular}
\end{ruledtabular}
\begin{tabbing}
$^{\mathrm{a}}$ Literature values of $\langle r^{-1} \rangle$ for first three states are: 0.579336, 0.572186, 0.562375 \cite{roy04}. \\
$^{\mathrm{b}}$ Literature values of $\langle r^1 \rangle$ for first three states are: 1.888604, 2.203514, 2.495625 \cite{roy04}. 
\end{tabbing}
\end{table}
\endgroup
Once the accuracy and reliability is established, next in Table III, we report first 6 states of a charged harmonic oscillator 
belonging to angular quantum number $\ell=0$. All these states are obtainable from an $r_{max}=20$ a.u. A broad range
of the coupling parameter, \emph{viz.,} $\lambda= \pm 0.001, \pm 0.01, \pm 0.1, \pm 1, \pm 10$ is considered, covering a wide interaction
region. For $\lambda= \pm 0.001, \pm 0.1$ and $\pm 10$, the first three states have been calculated before through a generalized 
pseudospectral method \cite{roy04}. Current energies obtained from imaginary-time evolution technique are in quite good agreement with 
these literature values, quoted here in parentheses. While the current results do not reach the precision of \cite{roy04} within our
present implementation, these are certainly still very good and almost for all practical purposes, sufficiently accurate. No other 
results are available at this time for other states. In addition, for each 
of these states, the position expectation values $\langle r^{-1} \rangle$ and $\langle r \rangle$ are given as well.

\begingroup
\squeezetable
\begin{table}
\caption {\label{tab:table4} First six lowest eigenvalues (a.u.) of charged harmonic oscillator corresponding to $l=0$, for 
two values of $\lambda$, with variations in grid. $N_r$ implies number of radial points.} 
\begin{ruledtabular}
\begin{tabular}{cccccc}
$\lambda$  & $n $  & $N_r=501$     & $N_r=1001$      &  $N_r = 2001$ &  $N_r = 5001$  \\
\hline
 $-0.001$  &  0    &   1.4994081   &     1.4994275   &    1.4994337     &   1.4994357     \\
           &  1    &   3.4994911   &     3.4995136   &    3.4995271     &   3.4995298     \\ 
           &  2    &   5.4995407   &     5.4995717   &    5.4995793     &   5.4995815     \\  
           &  3    &   7.4995801   &     7.4996020   &    7.4996125     &   7.4996161     \\  
           &  4    &   9.4996278   &     9.4996374   &    9.4996400     &   9.4996416     \\    
           &  5    &  11.4996001   &    11.4996553   &   11.4996598     &  11.4996616     \\
0.01       &  0    &   5.2887423   &    5.2887418    &    5.2887417     &   5.2887417     \\ 
           &  1    &   7.0754429   &    7.0754398    &    7.0754394     &   7.0754394     \\
           &  2    &   8.8981276   &    8.8981178    &    8.8981166     &   8.8981164     \\  
           &  3    &  10.7479984   &   10.7479717    &   10.7479674     &  10.7479670     \\
           &  4    &  12.6188822   &   12.6188020    &   12.6187939     &  12.6187932     \\
           &  5    &  14.5063045   &   14.5061641    &   14.5061505     &  14.5061493     \\ 
\end{tabular}
\end{ruledtabular}
\end{table}
\endgroup

At this point, Table IV gives a comparison of energies obtained in various grids. For this purpose, two $\lambda$ values of 
$-$0.001 and 10 of 
the charged harmonic oscillator are selected. All six eigenvalues are considered for four $N_r$ values, \emph{viz.,} 501, 1001, 2001 
5001, keeping the initial guess same in all occasions. It is clearly seen that, even the smallest grid produces results accurate
up to fourth place of decimal except the highest state corresponding to $\lambda=10$. For all the states, however, the results 
improve with successive increase in $N_r$, i.e., a denser grid is needed. The simulation box was roughly
15 a.u. As already mentioned, while such tests are not undertaken for all the potentials under study, it has, however, been 
verified that, the last grid $N_r=5001$ is sufficient to completely reproduce all the eigenvalues of previous table for these two
$\lambda$ values. Nevertheless, from the experience of these two cases, we believe this could be equally true for the other potential 
sets in the table as well.   

As a last example, Table V reports ground-state energies of SHO for two values of $\alpha=4$ (left) and 6 (right) for small as well
as large $\lambda$s. Note that in the last 
three decades, there has been significant interest in this system due to its many fascinating characteristics. One distinctive feature
of such a potential is that once the perturbation $\lambda |r|^{-\alpha}$ is turned on, it is impossible to \emph{completely turn off} the
interaction. Also, in the region of $\alpha \geq 5/2$, it exhibits \emph{super-singularity}. For many other facets of this potential, 
the reader is referred to the following references \cite{detwiler75,killingbeck82,fernandez91,navarro94,torres92,buendia95,handy96,
gomez10,roy04}, Both the $\alpha$ values considered 
can lead to super-singularity; these have been studied by numerous analytic, semi-analytic as well as numerical methodologies. Some of 
these literature results are
given here for comparison. It is seen that the present methodology offers results which are in good agreement with these. The most 
accurate results are those from analytic continuation method \cite{buendia95} and generalized pseudospectral method \cite{roy04}. 
The present energies are not superior to these, but still are excellent and evidently better than many other reference values. 

\begingroup
\squeezetable
\begin{table} 
\caption {\label{tab:table5}Calculated ground-state energies E (in a.u.) of the SHO with $\alpha=4$ and 6 for several values of $\lambda.$ 
The literature results are divided by a 2 factor. PR signifies Present Result.} 
\begin{ruledtabular}
\begin{tabular}{cllll}
$\lambda$ & \multicolumn{2}{c}{Energy ($\alpha=4$)} & 
\multicolumn{2}{c}{Energy ($\alpha=6$)} \\ 
\cline{2-3} \cline{4-5}
      & PR          & Literature                                                                 & PR              & Literature \\   \hline
0.001 & 1.53438158  & 1.53438158\footnotemark[1]$^,$\footnotemark[2], 1.534385\footnotemark[3] 
      & 1.63992791  & 1.63992791\footnotemark[1]$^,$\footnotemark[2]  \\
0.005 & 1.57417615  & 1.57417615\footnotemark[2]$^,$\footnotemark[4],                  
      & 1.71144209  & 1.71144209\footnotemark[2],1.71144208\footnotemark[4], \\
      &             & 1.574175\footnotemark[5], 1.574195\footnotemark[6]                         
      &             & 1.71144\footnotemark[5],1.71151\footnotemark[6] \\
0.01  & 1.60253374  & 1.60253374\footnotemark[1]$^,$\footnotemark[2], 1.60254\footnotemark[3], 
      & 1.75272613  & 1.75272613\footnotemark[1]$^,$\footnotemark[2],1.752726195\footnotemark[4], \\ 
      &             & 1.60253374\footnotemark[4],1.602535\footnotemark[5]$^,$\footnotemark[7],1.602635\footnotemark[6]   
      &             & 1.752725\footnotemark[5],1.75287\footnotemark[6],1.7527265\footnotemark[7]       \\
0.05  & 1.71258069  & 1.71258069\footnotemark[2]         
      & 1.88277010  & 1.88277010\footnotemark[2]            \\
0.1   & 1.78777599  & 1.78777599\footnotemark[1]$^,$\footnotemark[2],1.787785\footnotemark[3],1.787775\footnotemark[7] 
      & 1.95783261  & 1.95783261\footnotemark[2]            \\ 
0.5   & 2.06529243  & 2.06529243\footnotemark[2]            
      & 2.19395453  & 2.19395453\footnotemark[2]            \\
1     & 2.24708899  & 2.24708899\footnotemark[1]$^,$\footnotemark[2],2.24709\footnotemark[3]$^,$\footnotemark[7]  
      & 2.32996998  & 2.32996998\footnotemark[1]$^,$\footnotemark[2],2.329970\footnotemark[7]  \\ 
5     & 2.89222177  & 2.89222177\footnotemark[2],2.89222\footnotemark[7] 
      & 2.75657950  & 2.75657950\footnotemark[2],2.7565795\footnotemark[7]               \\
10    & 3.30331125  & 3.30331125\footnotemark[1]$^,$\footnotemark[2]$^,$\footnotemark[8],3.30331\footnotemark[3]$^,$\footnotemark[7]  
      & 3.00160451  & 3.00160451\footnotemark[1]$^,$\footnotemark[2],3.0016045\footnotemark[7],3.00160451\footnotemark[8] \\  
50    & 4.73277787  & 4.73277787\footnotemark[2]            
      & 3.76776072  & 3.76776072\footnotemark[2]        \\
100   & 5.63254021  & 5.63254021\footnotemark[1]$^,$\footnotemark[2],5.63254\footnotemark[3],5.6325402\footnotemark[8]  
      & 4.20667914  & 4.20667914\footnotemark[2]$^,$\footnotemark[8]    \\ 
500   & 8.73793385  & 8.73793385\footnotemark[2]     
      & 5.57607711  & 5.57607711\footnotemark[2]             \\
1000  & 10.6847312  & 10.6847312\footnotemark[1]$^,$\footnotemark[2]$^,$\footnotemark[8],10.68473\footnotemark[3]
      & 6.35930853  & 6.35930853\footnotemark[2]   \\
\end{tabular}
\begin{tabbing}
$^{\mathrm{a}}${Ref.~\cite{buendia95}.} \hspace{15pt} \=  
$^{\mathrm{b}}${Ref.~\cite{roy04}.} \hspace{15pt} \=
$^{\mathrm{c}}${Ref.~\cite{navarro94}.} \hspace{15pt} \=
$^{\mathrm{d}}${Ref.~\cite{torres92}.} \hspace{15pt} \= 
$^{\mathrm{e}}${Ref.~\cite{killingbeck82}.} \hspace{15pt} \=
$^{\mathrm{f}}${Ref.~\cite{detwiler75}.} \hspace{15pt} \=
$^{\mathrm{g}}${Ref.~\cite{fernandez91}.} \hspace{15pt} \=
$^{\mathrm{h}}${Ref.~\cite{handy96}.} 
\end{tabbing}
\end{ruledtabular}
\end{table}
\endgroup

Finally, to show the quality of our wave functions obtained, we depict the radial distribution functions of charged harmonic oscillator 
in Fig.~1. Diagrams (a)--(e) in left panel correspond to the potential (a) (with $\alpha=1, \lambda=0.01$) and first four low-lying 
state densities respectively, with (b) referring to that of ground state. Similarly in (f)--(j) in right panel, plots for potential (f) 
(with $\alpha-1, 
\lambda=-10)$ and four lowest states are displayed, with (g) identifying the lowest state. In both cases, density plots for all
states are given in same scale of radial distance. They both carry the signatures of acceptable eigenfunctions with number of nodes
increasing with state index. It is seen that, in the right side, peak height decreases to a greater extent as one goes to higher 
excitations, compared to the potential in left side.  


\begin{figure}
\begin{minipage}[c]{0.40\textwidth}\centering
\includegraphics[scale=0.26]{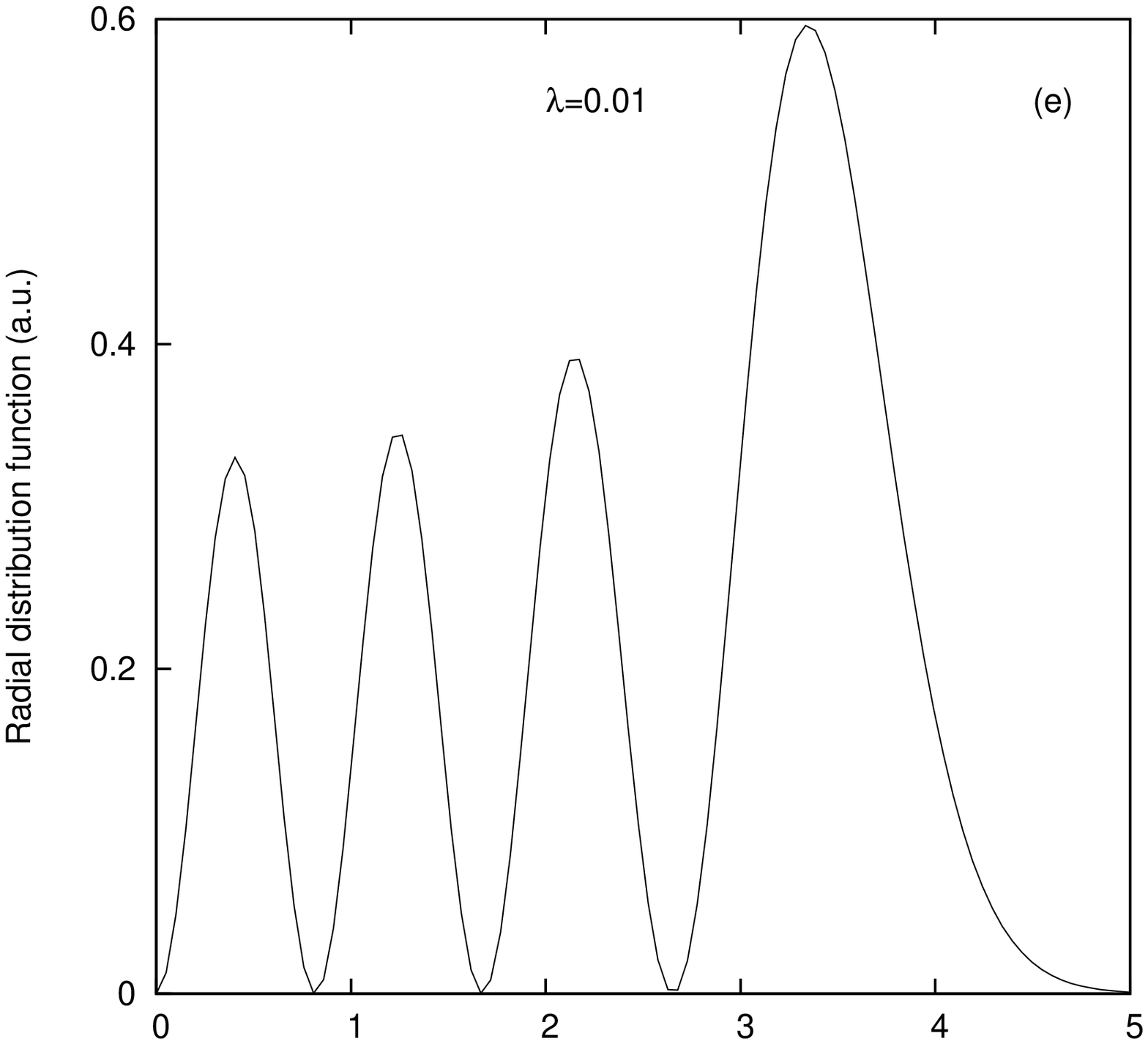}
\end{minipage}
\hspace{0.15in}
\begin{minipage}[c]{0.40\textwidth}\centering
\includegraphics[scale=0.26]{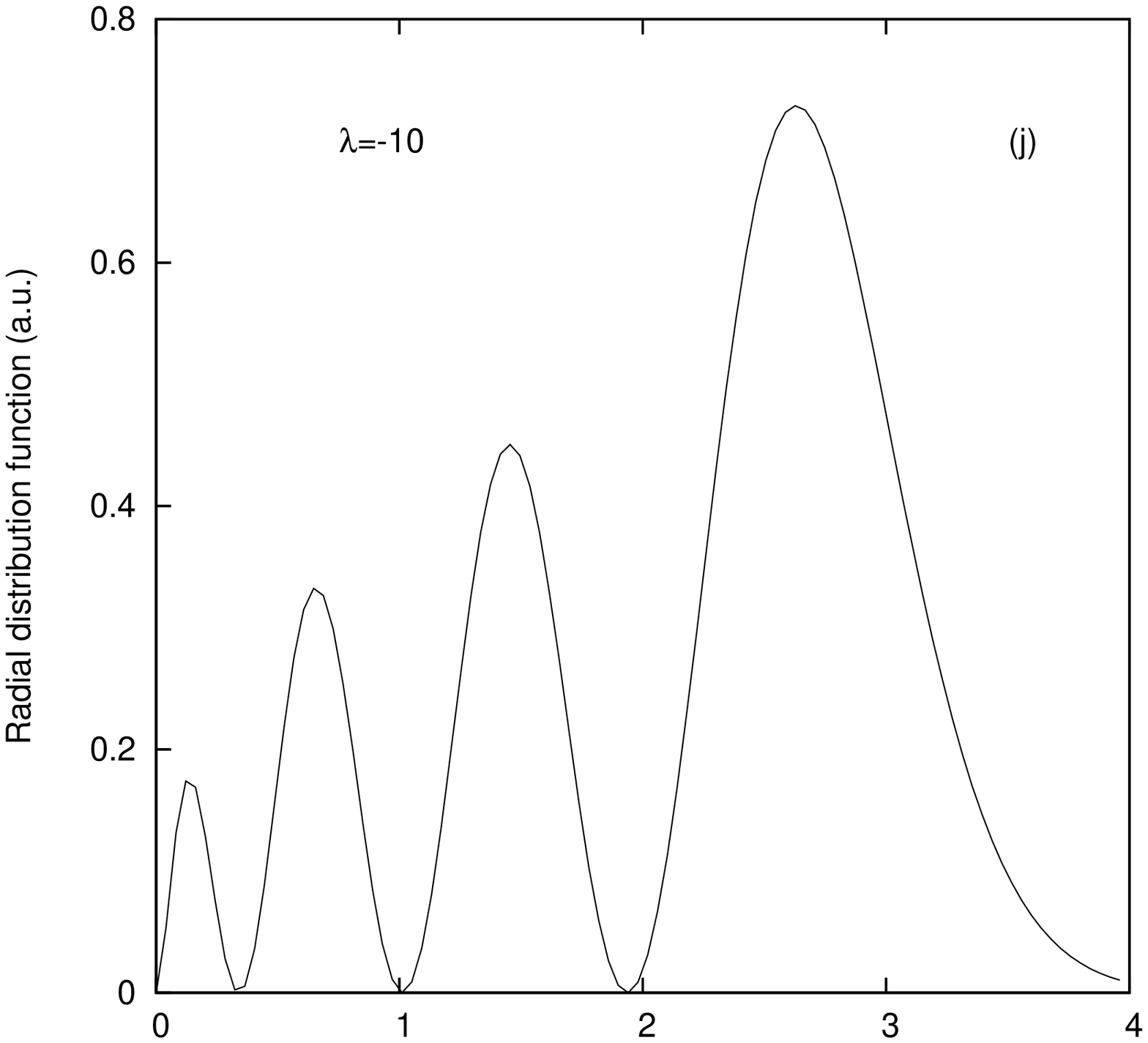}
\end{minipage}
\\
\begin{minipage}[b]{0.40\textwidth}\centering
\includegraphics[scale=0.26]{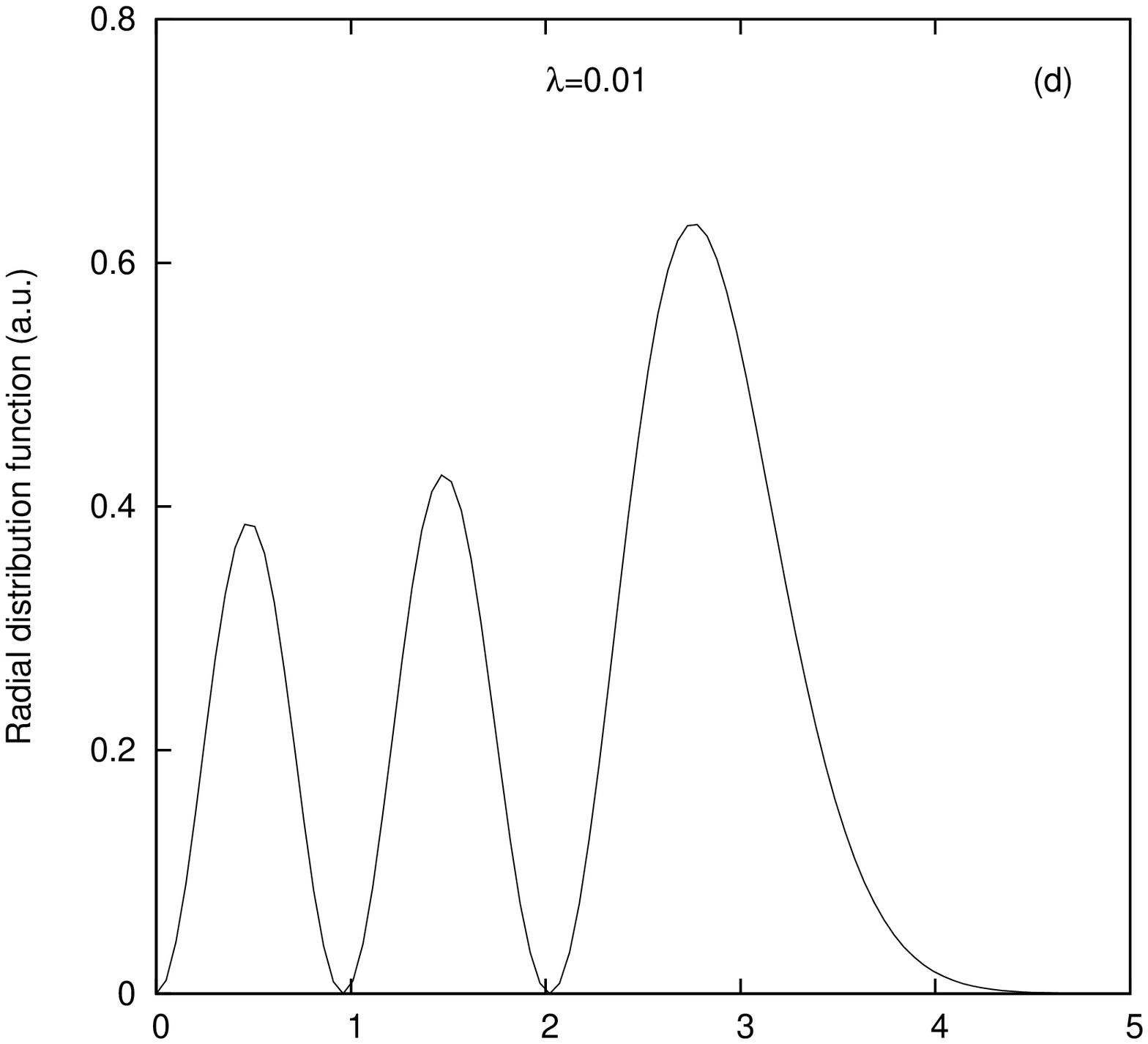}
\end{minipage}
\hspace{0.15in}
\begin{minipage}[b]{0.40\textwidth}\centering
\includegraphics[scale=0.26]{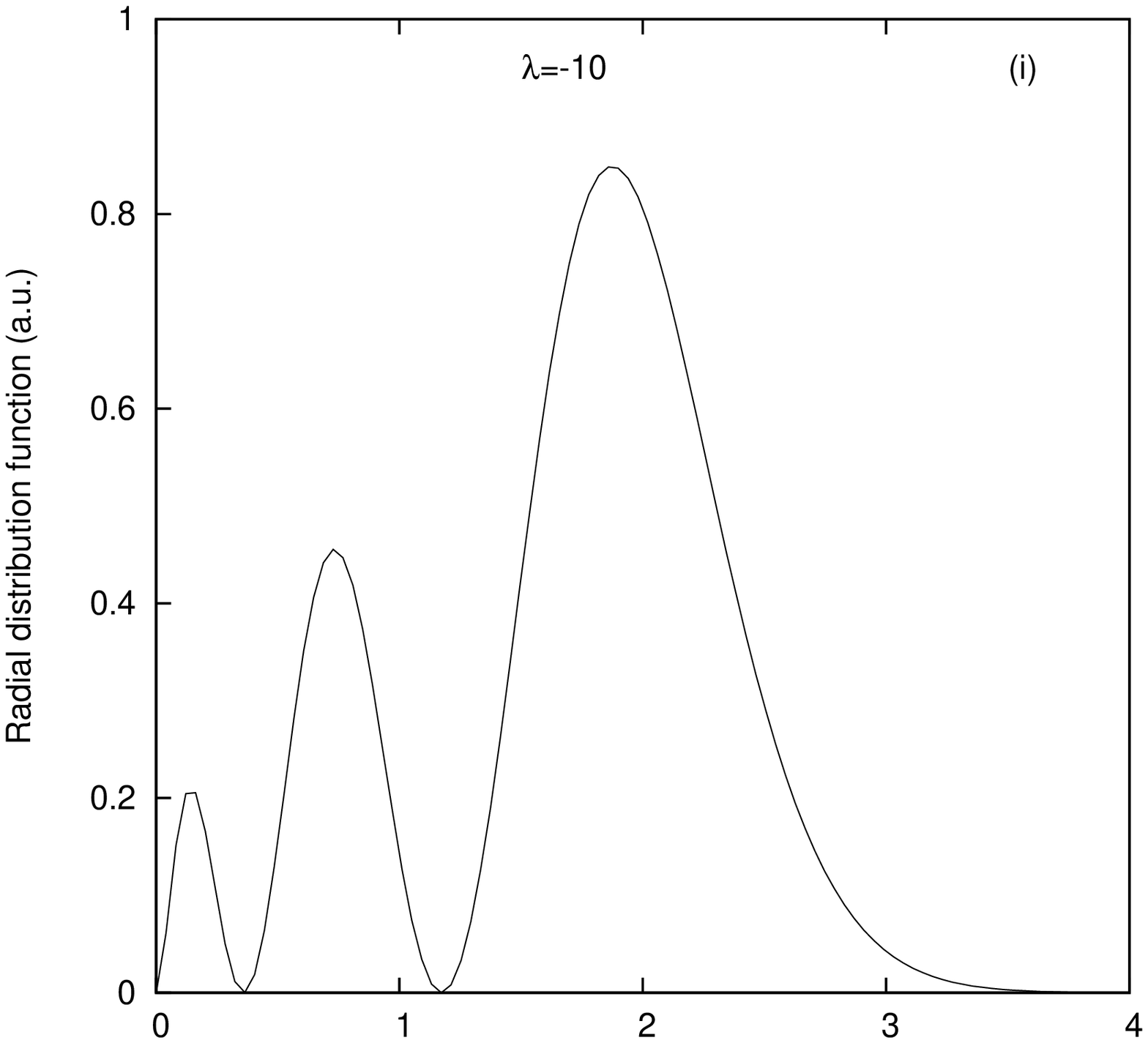}
\end{minipage}
\\
\begin{minipage}[b]{0.40\textwidth}\centering
\includegraphics[scale=0.26]{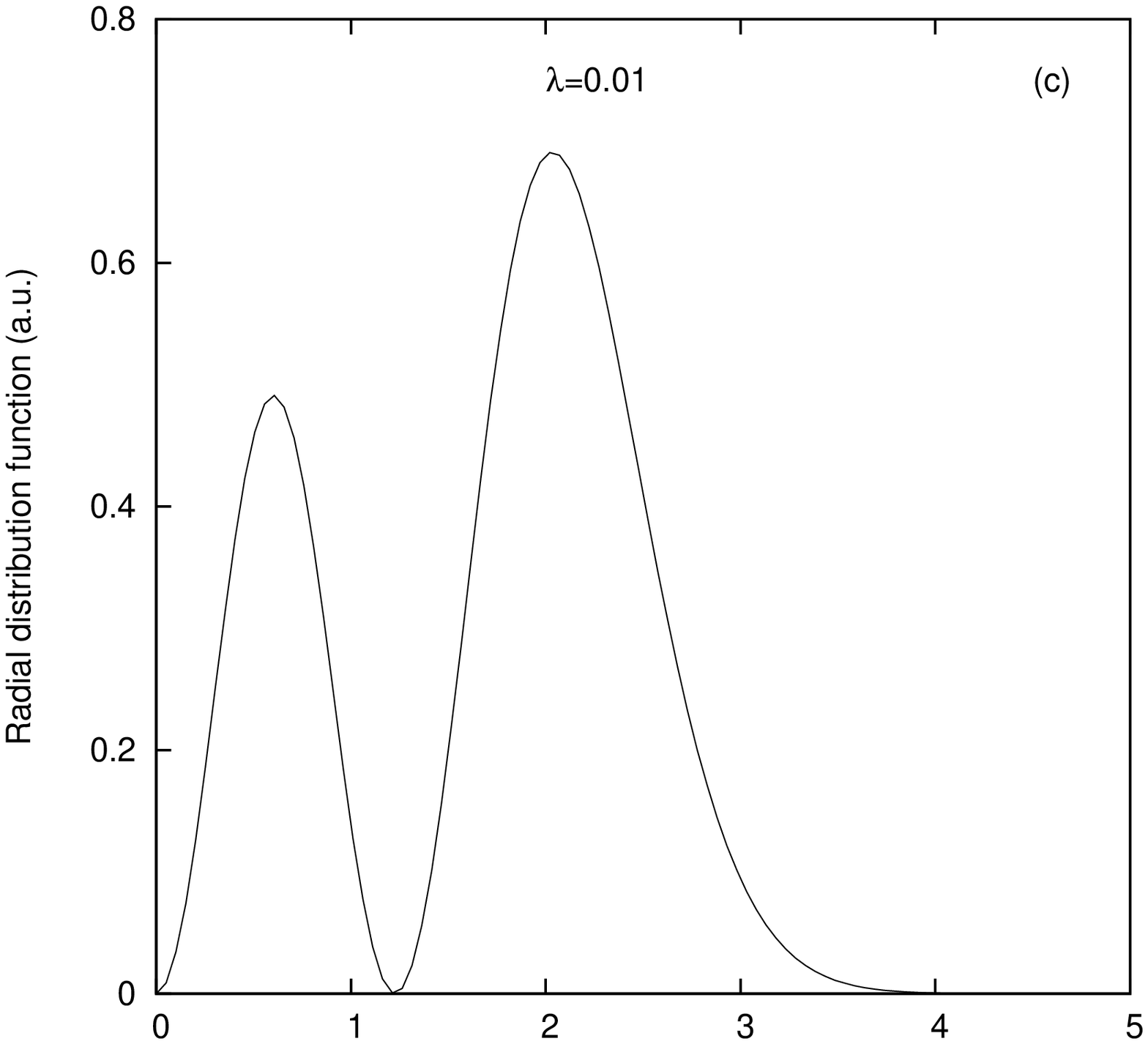}
\end{minipage}
\hspace{0.15in}
\begin{minipage}[b]{0.40\textwidth}\centering
\includegraphics[scale=0.26]{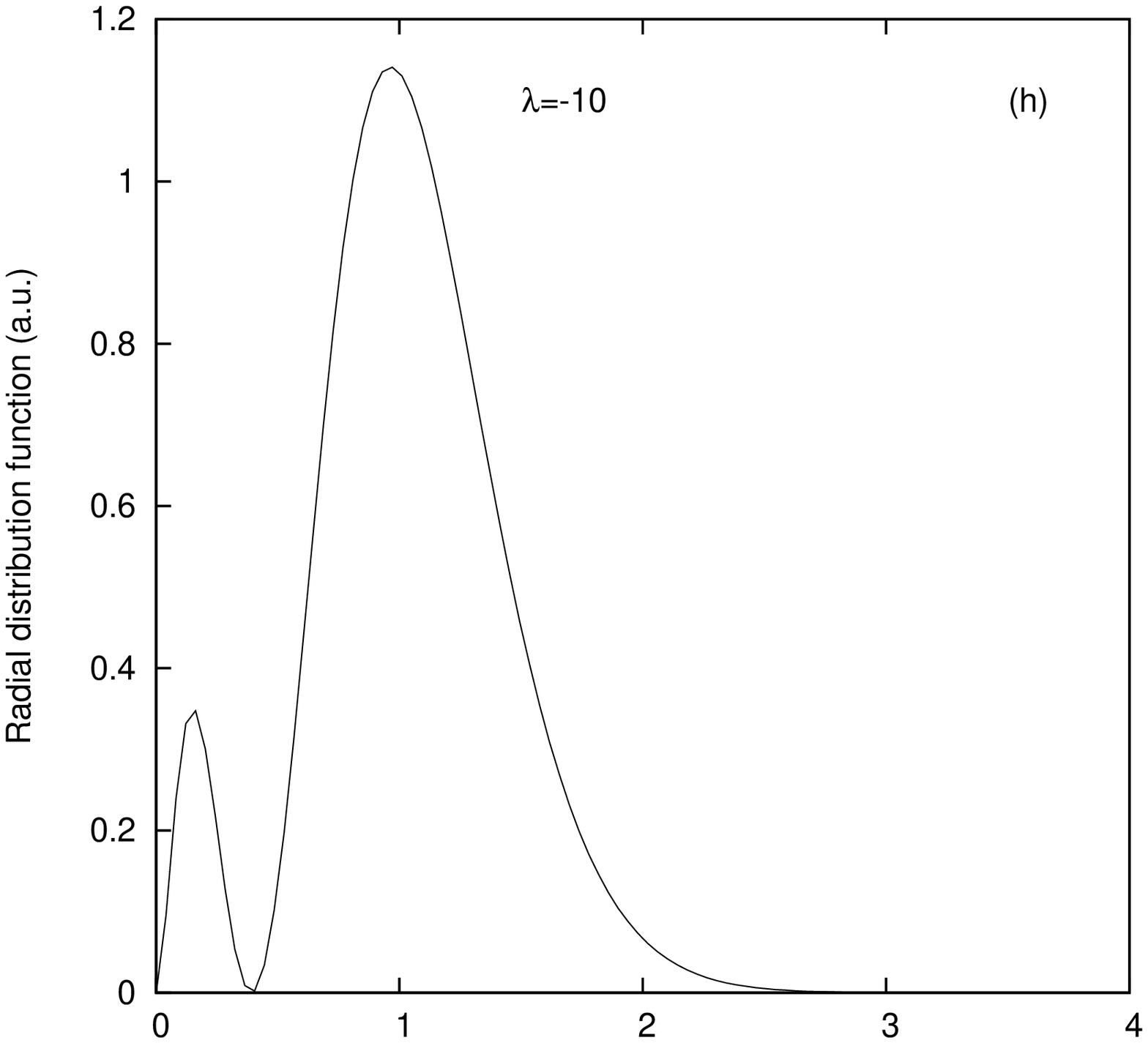}
\end{minipage}
\\
\begin{minipage}[b]{0.40\textwidth}\centering
\includegraphics[scale=0.26]{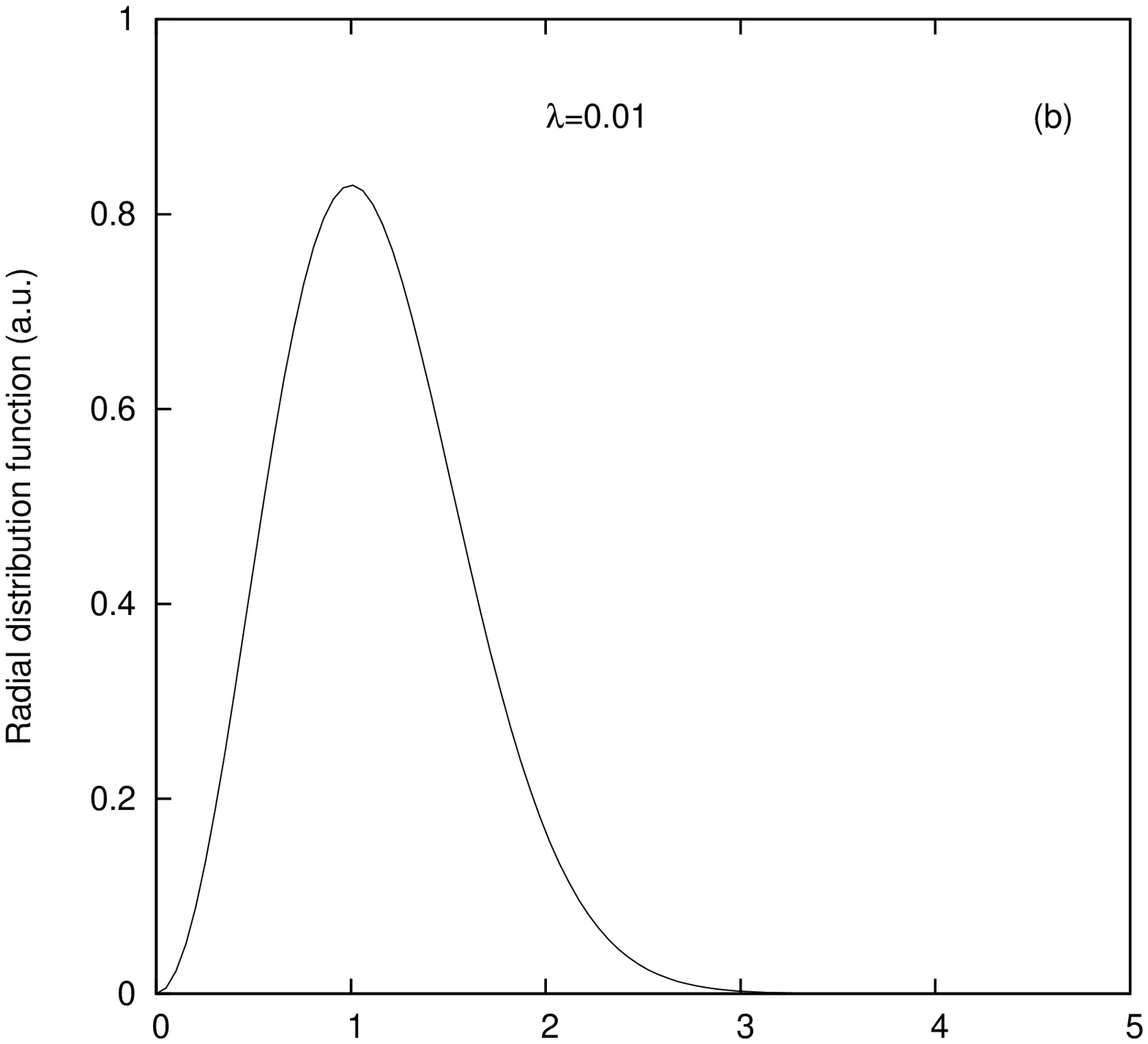}
\end{minipage}
\hspace{0.15in}
\begin{minipage}[b]{0.40\textwidth}\centering
\includegraphics[scale=0.26]{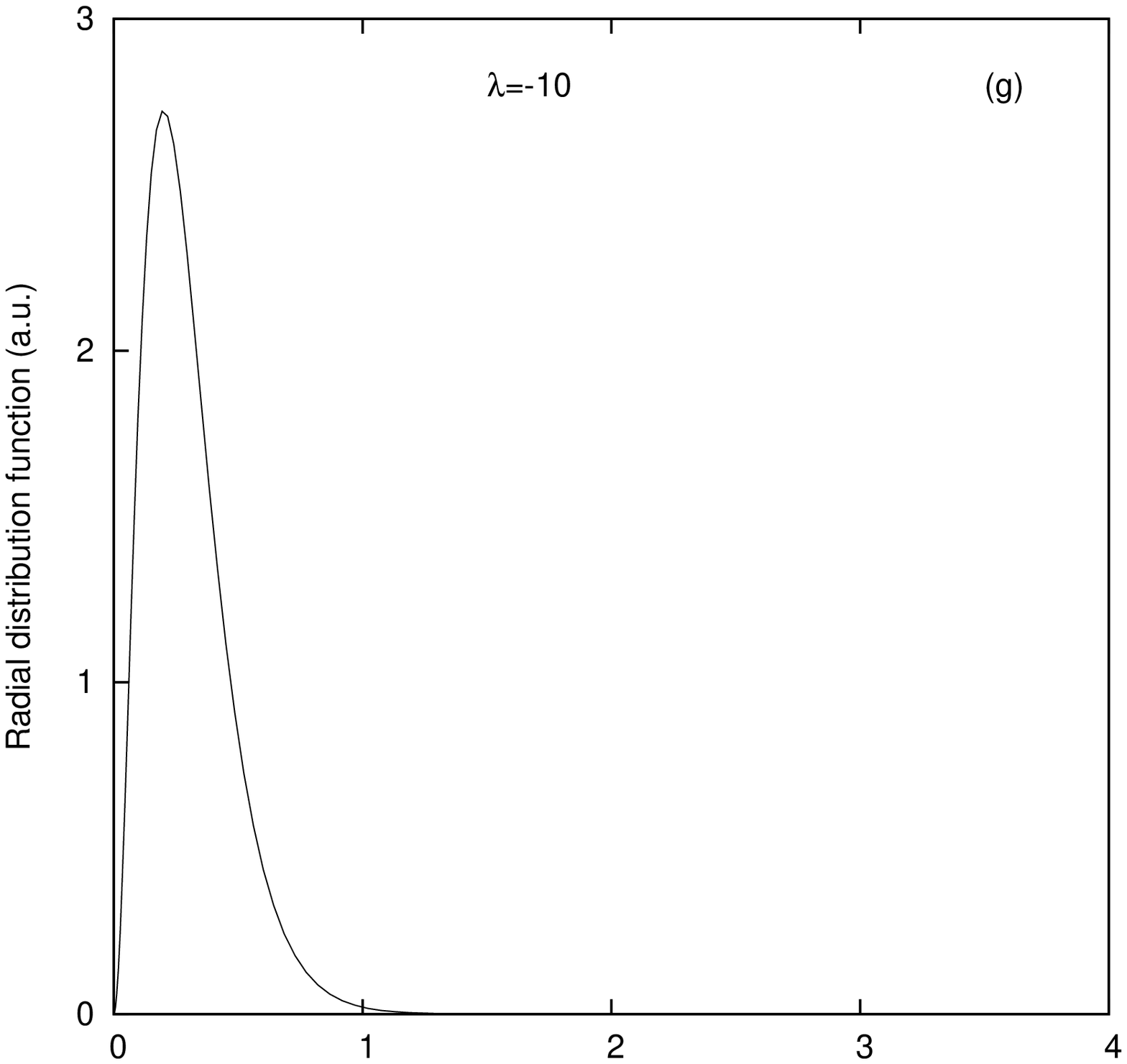}
\end{minipage}
\\
\begin{minipage}[b]{0.40\textwidth}\centering
\includegraphics[scale=0.26]{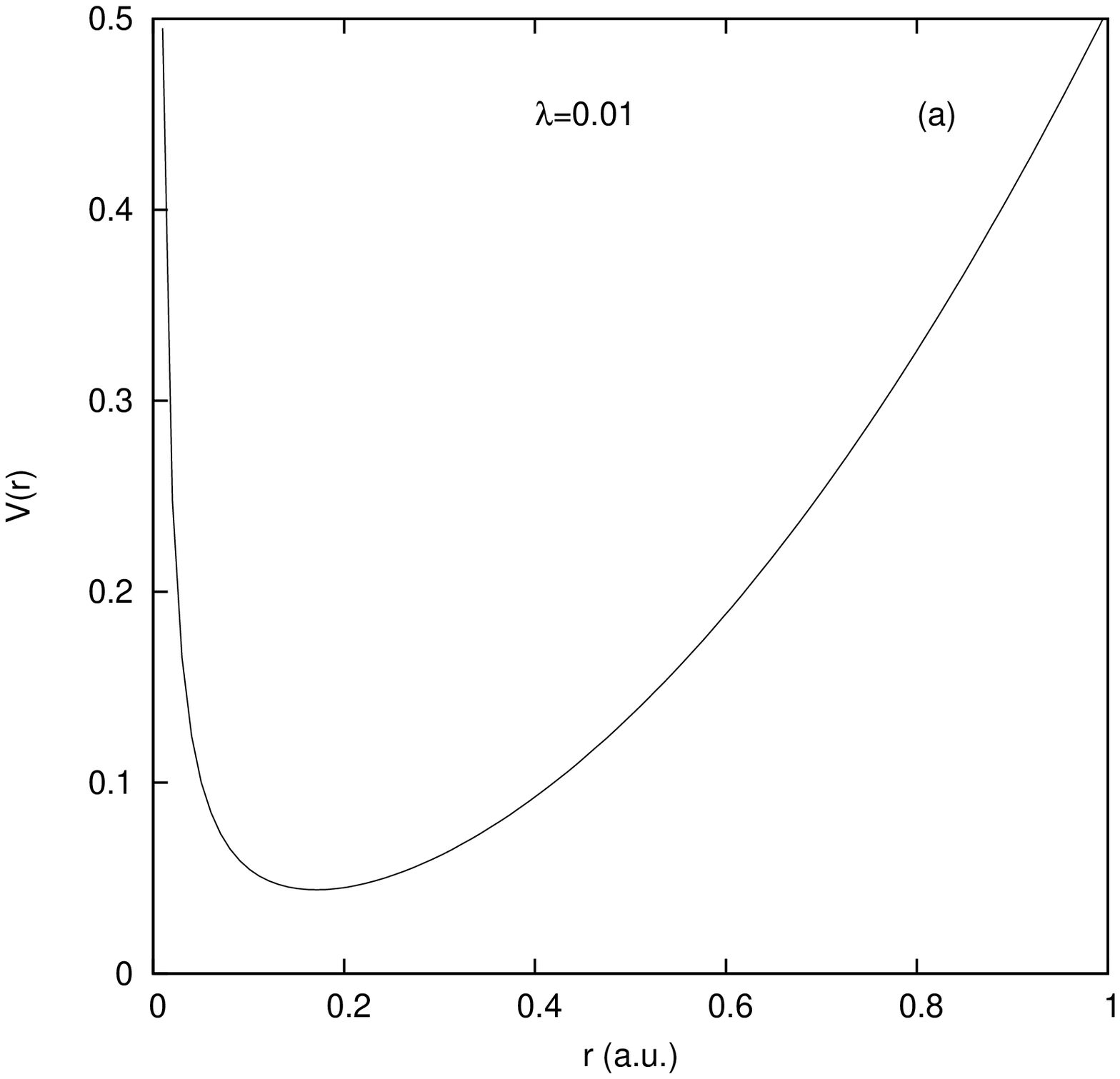}
\end{minipage}
\hspace{0.15in}
\begin{minipage}[b]{0.40\textwidth}\centering
\includegraphics[scale=0.26]{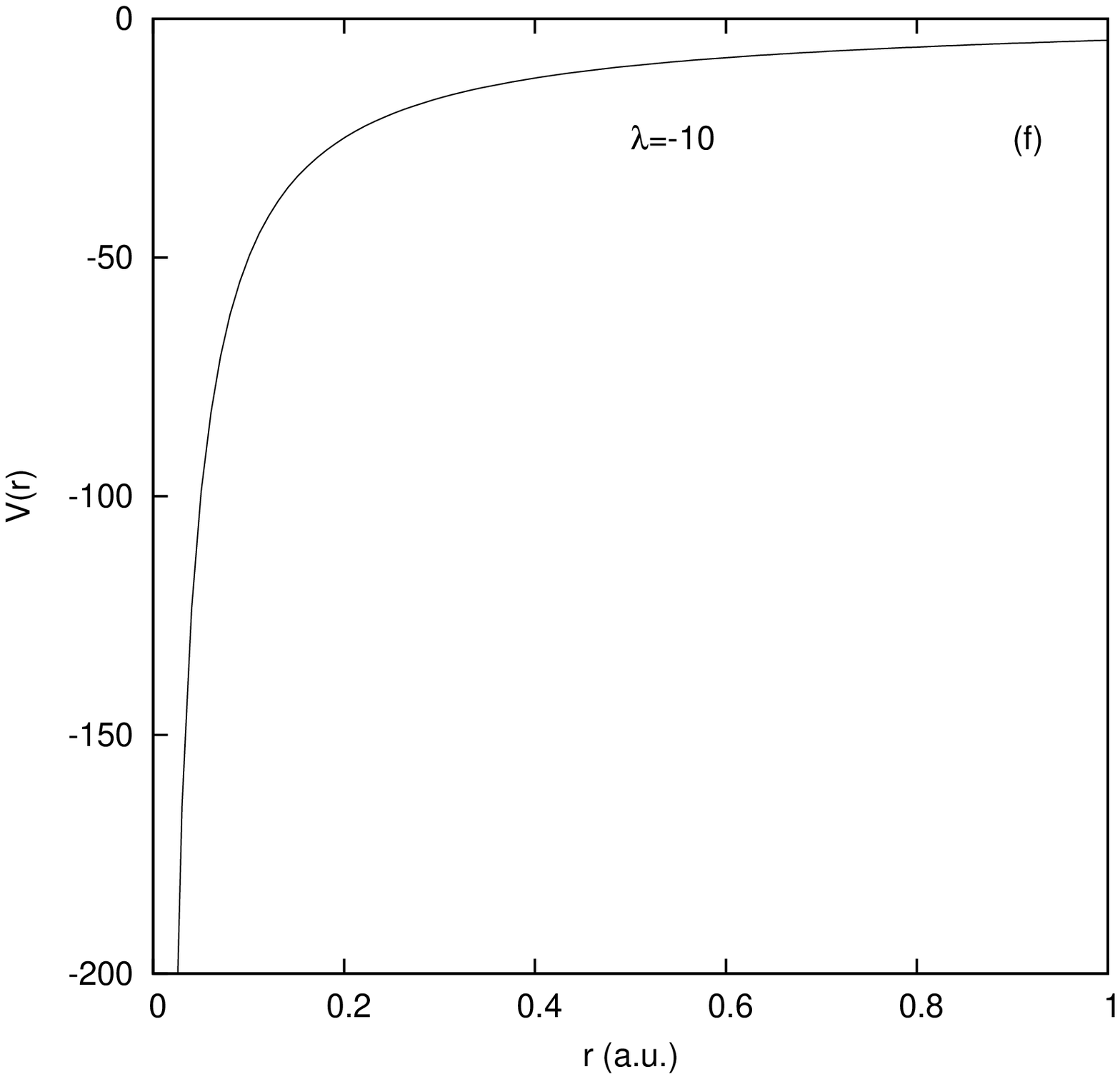}
\end{minipage}
\caption{The radial probability distribution functions of charged harmonic oscillators. Left and right panels corresponds to 
$\lambda=0.01$ and $-$10. The two potentials are shown in $\{(a), (f)\}$, while $\{(b), (g)\}$; $\{(c), (h)\}$; $\{(d), (i)\}$; 
$\{(e), (j)\}$ refer to the densities of ground, first, second and third excited states respectively, corresponding to 
$\ell=0$.}\label{fig:fig}
\end{figure}

A few words may be devoted to the initial trial function. Most of our calculations were performed with Gaussian-type functions 
as starting guess to launch the computations. However, several other sample guessed (including some wild) functions were tried to test 
the efficiency of this formalism. In such cases, the effective computation time required to achieve convergence of desired accuracy 
varies with initial guess, as more and more iterations are required. Generally, it 
was found that, keeping all things unchanged, during iterative process, mixing diffusion function with that from previous
time step by a certain percentage (we employed a 50:50 mixture) increased the rate of convergence. Accuracy of the present method
depends on density of the grid and propagation time. The degeneracy in case of symmetric and non-symmetric 2D double-well oscillators
\cite{roy05} as well as pseudo-degeneracy in 1D double wells \cite{roy02, gupta02} have been well represented by this method. 
It is conceivable that convergence and accuracy 
of our results could be further improved by choosing different spatial grid, more appropriate and suitable initial wave functions, 
higher-order finite difference schemes as well as higher precision computation, some of which may be taken up later.   
 
\section{conclusion}
Energy eigenvalues, select position expectation values and probability densities of 3D spherically symmetric potentials are obtained
accurately and efficiently by means of an imaginary time evolution method in conjunction with minimization of an energy expectation
value. Numerical propagation of the resulting diffusion equation eventually hits ground state and ensuring orthogonalization
to lower states, leads to excited states in a sequential manner.
Comparison with available literature data reveals that good-quality, meaningful results could be produced in all the occasions
concerned. Thus it could pose a viable alternative to the existing methodologies available for such systems. This is illustrated
for a variety of systems, such as quantum harmonic oscillator, Morse potential, charged harmonic oscillator and spiked harmonic 
oscillator. The present work, as such, remains valid for spherically symmetric potentials. And therefore may not be directly applicable to 
situations, where the same is not possible, such as that in \cite{fertig00}. Another disconcerting feature of the method lies in the 
fact that for excited state calculations it must maintain orthogonality requirement with respect to all other lower states of same 
symmetry. That means, unless all the lower-state wave functions are properly converged in the active grid space, accurate results
would be difficult for excited states. 
Future applications of the method to non-zero angular states, as well as other interacting potentials of physical interest, such as 
molecular, atomic, screening, power-law, rational etc., and quantum confinement studies may further consolidate the success 
of this approach. 

\section{acknowledgment} It is gratefully acknowledged. It is a pleasure to thank Mr.~Siladitya Jana for supplying some of the
references.

\end{document}